\documentclass[prl,twocolumn,showpacs, superscriptaddress]{revtex4}

\usepackage{amsmath}
\usepackage{amssymb}
\usepackage{amsthm}
\usepackage{graphicx}

\newcommand{\tr}{\operatorname{tr}}
\newcommand{\uinvnorm}{|\kern-2pt|\kern-2pt|}

\begin{document}
\bibliographystyle{apsrev}

\title{Achievable Qubit Rates for Quantum Information Wires}

\author{Hulya Yadsan-Appleby}
\email[]{h.yadsan-appleby@ucl.ac.uk} \affiliation{Department of Physics \& Astronomy, University College London, Gower Street, London WC1E 6BT}
\author{Tobias J.\ Osborne}
\email[]{tobias.osborne@itp.uni-hannover.de} \affiliation{Institut f{\"u}r Theoretische Physik, Appelstra{\ss}e 2, 30167 Hannover, Germany}
\date{25 October 2010}

\begin{abstract}
	Suppose Alice and Bob have access to two separated regions, respectively, of a system of electrons moving in the presence of a regular one-dimensional lattice of binding atoms. We consider the problem of communicating as much quantum information, as measured by the \emph{qubit rate}, through this \emph{quantum information wire} as possible. We describe a protocol whereby Alice and Bob can achieve a qubit rate for these systems which is proportional to $N^{-\frac{1}{3}}$ qubits per unit time, where $N$ is the number of lattice sites. Our protocol also functions equally in the presence of interactions modelled via the $t-J$ and Hubbard models.  
\end{abstract}

\pacs{03.67.Hk, 05.50.+q, 32.80.Lg}

\maketitle
One of the key requirements for a functioning quantum information processor is the ability to transport quantum information from one location to another (see, e.g., \cite{divincenzo:2000a} for further details.) Hence, finding physical systems usable as error-tolerant quantum communication channels has become a priority. Recently a class of physical systems realising such channels, namely strongly interacting quantum 
spin systems, has been investigated \cite{bose:2003a, christandl:2004a, osborne:2004a, haselgrove:2005a, kay:2010a, murphy:2010a, paganilli:2009a, schirmer:2009a, yung:2006a, yang:2006a, bose:2007a, lewenstein:2007a, ciaramicoli:2007a}. Under fairly general conditions it has been shown that such \emph{quantum information wires} can communicate quantum information with arbitrarily high fidelity. 
 
Generally speaking, the proposals developed so far explore hilbert space rather inefficiently: typically, once a signal has been placed in the wire, Alice must wait until Bob has successfully removed the signal before another signal can be sent. In other words, the \emph{qubit rate} $Q$ --- the number of qubits which can be successfully communicated per unit time --- for these protocols scales roughly inversely with the length of the system, i.e., $Q \sim N^{-1}$, where $N$ is the number of lattice sites. So far there has only been one proposal \cite{kay:2010a,kay:2009a,kay:2006a} exploring qubit rates beyond this: there a protocol for an \emph{engineered} chain is described which achieves, after taking energy normalisation into account, a qubit rate $Q\sim N^{-\frac12}$.

In this Letter we aim to improve this situation in two ways: firstly, we study naturally occurring one-dimensional systems of particles moving in the presence of a regular lattice of binding atoms, and secondly describe a protocol achieving a qubit rate $Q\sim N^{-\frac13}$. Our results do not require that the interactions between the particles are engineered, however, we do assume that the sender and receiver each have good control of at most one lattice site each.

We initially study free fermions moving in the presence of a regular one-dimensional lattice of $N$ atoms. The number $N \gg 1$ is considered to be a large parameter which is divisible by $4$ for convenience. We denote by $\{a_j\}_{j=1}^N$ the operators annihilating a fermion at site $j$; the operators $\{a_j\}_{j=1}^N$ obey the canonical anticommutation relations $\{a_j, a_k^\dag\} = \delta_{jk}$. Thus our hilbert space is \emph{fock space}, and has an orthonormal basis given by the \emph{occupation number basis}: $|n_1, n_2, \ldots, n_N\rangle \equiv (a_1^\dag)^{n_1}(a_2^\dag)^{n_2}\cdots (a_N^\dag)^{n_N}|\Omega\rangle$, where $n_j\in\{0,1\}$ are the number of electrons located at site $j$, and we identify the vacuum state $|\Omega\rangle \equiv |0,0, \ldots, 0\rangle$. We initially model the lattice geometry using the \emph{tight-binding model} \cite{ashcroft:1976a}, describing free electrons hopping on a regular lattice of $N$ sites:
\begin{equation}\label{eq:tightbinding}
	H = \sum_{j=1}^{N} a_j^\dag a_{j+1} +\text{h.c.},
\end{equation}
where we identify site $(N+1)$ with site $1$, i.e., we assume periodic boundary conditions. The periodic geometry is a theoretical device and we later argue that our results extend straightforwardly to the more realistic chain setting. We also argue later that our protocol functions equally well in the presence of interactions modelled by the $t-J$ model and the Hubbard model. 

We define the length of the system to be $1$. Therefore, the \emph{single-particle} quantum state $|j\rangle \equiv |0,0, \ldots, 1_j, \ldots, 0\rangle$ means that there is a particle sitting with probability one at \emph{position} $x_j = j/N$ in the lattice. We refer to the subspace $\mathcal{H}_S$ spanned by $|j\rangle$ as the \emph{single-particle subspace}. When we refer to $|j\rangle$ we say that the particle is at \emph{site} $j$ corresponding to \emph{physical position} $x_j$.

As the number of lattice sites increases, and if the quantum state of the system is the discretisation of a sufficiently smooth wavefunction, the dynamics will be equivalent to that of a free particle on a circle. (See the supplementary material for further details). The consequence of this is that gaussian-modulated wavepackets with wavenumber $k$ propagate at some \emph{group velocity} $v(k)$ with essentially no change in shape. 

In our protocol Alice wants to communicate a sequence of (possibly unknown) qubit states $|\psi_\alpha\rangle$, $\alpha=1, 2, \ldots, M$, to Bob. Alice and Bob are each allowed access to a region $R_A = \{1, 2, \ldots, \nu N^{\frac13}\}$, ($R_B = \{N/2, N/2+1, \ldots, N/2+\nu N^{\frac13}-1\}$, respectively) , of $\nu N^{\frac{1}{3}}$ sites, where $\nu$ is a constant to be chosen later, at opposite ends of the ring. (We later argue that Alice and Bob only need access to one site within their regions, respectively.) Initially the system is prepared in the vacuum state $|\Omega\rangle$. Alice and Bob each have access to $M$ ancilla qubits labelled ${A_\alpha}$,  (respectively, ${B_\alpha}$): the qubits ${A_\alpha}$ are initialised in a product of the $M$ qubit states $|\psi_\alpha\rangle$ and ${B_\alpha}$ are initialised in a product of some convenient fiducial state $|0\rangle$. Thus the initial state of the system before the protocol begins is $|\psi_1\rangle_{A_1}\cdots |\psi_M\rangle_{A_M} |\Omega\rangle |0\rangle_{B_1}\cdots |0\rangle_{B_M}$.

To accomplish the communication task, Alice performs some \emph{encoding operation} $U_1$ coupling her region $R_A$ and the qubit $A_1$ containing the first message $|\psi_1\rangle$. The system is now in the state $U_1|\psi_1\rangle|\Omega\rangle$. (From now on, when discussing the state of the system we suppress mention of the qubits which haven't yet interacted with the lattice.) The system evolves as $|\Psi(t)\rangle = e^{-iHt}U_1|\psi_1\rangle|\Omega\rangle$. After a time $T$, where  $T$ depends only on $H$ and $\nu N^{\frac13}$, Bob performs a \emph{decoding operation} $V_1$ on his addressable sites $R_B$ in order to decode or \emph{refocus}  the communicated state into the ancilla qubit $B_1$. The state of the system at the moment Bob has applied $V_1$ is $V_1 e^{-iHT} U_1|\psi_1\rangle|\Omega\rangle |0\rangle$. The protocol is deemed to succeed when the \emph{average fidelity} of the decoded state $\rho_{1} = \tr_{\widehat{B}_1}(V_1 e^{-iHT} U_1 |\psi_1\rangle\langle \psi_1| \otimes|\Omega\rangle\langle \Omega|\otimes  |0\rangle\langle 0| U_1^\dag e^{iHT} V_1^\dag)$, where $\tr_{\widehat{B}_j}$ denotes the partial trace over all the system+ancillas except for $B_j$, given by
\begin{equation}\label{eq:avgfid}
\mathcal{F}_1 \triangleq \frac{1}{4\pi}\int
d\Omega_1 \, \langle\psi_1|\rho_1|\psi_1\rangle,
\end{equation}
where $|\psi_1\rangle$ is the input state averaged over the Bloch sphere, is above some prespecified threshold value $1-\epsilon$. In the case where Alice wants to send $M$ signals we require that \emph{each} of the average fidelities $\mathcal{F}_\alpha$, of the $M$ messages is above $1-\epsilon$. We shall later obtain a lower bound for the average fidelity of the protocol we're about to describe.

The operations $U_\alpha$ are designed to simply \emph{swap} an unknown qubit state $|\psi_\alpha\rangle$ from register $\alpha$ into her accessibility region $R_A$; they are realised by applying the unitary operator
\begin{equation}
	U_\alpha = e^{-i\frac{\pi}{2}(\sigma^{+}_\alpha\sigma^{-}_\alpha gg^{\dag} + \sigma^{-}_\alpha\sigma^{+}_\alpha g^{\dag}g)}e^{i\frac{\pi}{2} (\sigma^{+}_\alpha g + \sigma^{-}_\alpha g^{\dag})},
\end{equation}
where $\sigma^{\pm}_\alpha = \frac{1}{2}(\sigma^x_\alpha \mp i\sigma^y_\alpha)$, $\{\sigma^x_\alpha, \sigma^y_\alpha, \sigma^z_\alpha\}$ are the Pauli sigma operators on qubit $A_\alpha$, and $g = \gamma\sum_{j \in R_A} e^{-\frac{(j-l)^2}{2\sigma^2}+2\pi ik
	j} a_j$,
with $\sigma$, $l$, and $k$ to be chosen later. The number $\gamma$ is chosen so as to ensure that $g$ obeys the canonical anticommutation relation $\{g, g^\dag\} = 1$, i.e.,$\{g, g^\dag\} = \gamma^2\sum_{j \in R_A} e^{-\frac{(j-l)^2}{\sigma^2}} = 1 $.
Note the useful identity
\begin{equation}\label{eq:Uidentity}
	U_\alpha = \mathbb{I} - \sigma^{+}_\alpha\sigma^{-}_\alpha gg^{\dag} - \sigma^{-}_\alpha\sigma^{+}_\alpha g^{\dag}g + \sigma^{+}_\alpha g + \sigma^{-}_\alpha g^{\dag}.
\end{equation}
The operator $g^\dag$ has been chosen so as to create a fermion in a discretisation of a gaussian envelope centred on $l = Nx_0$ in Alice's region with \emph{wavenumber} $k$. 

To understand how our encoding operation works we write $|\psi_1\rangle = c_1|0\rangle + d_1|1\rangle$. The transformation $|\psi_1\rangle |\Omega\rangle \mapsto U_1|\psi_1\rangle |\Omega\rangle$ is given by $U_1|\psi_1\rangle |\Omega\rangle = c_1|0\rangle|\Omega\rangle + d_1|0\rangle g^\dag|\Omega\rangle = |0\rangle (c_1\mathbb{I} + d_1g^\dag)|\Omega\rangle,$
it simply leaves the $|0\rangle$ state alone and flips the state $|1\rangle$ and creates a single fermion in a gaussian-modulated wavefunction in $R_A$. 

We can now understand the dynamics of the encoded state $|\Psi(0)\rangle = |0\rangle (c_1\mathbb{I} + d_1g^\dag)|\Omega\rangle$; we find that
\begin{equation}
	|\Psi(t)\rangle = |0\rangle (c_1\mathbb{I} + d_1g^\dag(t))|\Omega\rangle,
\end{equation}
where $g(t) = e^{-iHt}ge^{iHt}$. This is, in turn, given by the single-particle dynamics
$g(t) = \sum_{j,j'=1}^{N} g_j(0)[e^{it\triangle}]_{jj'} a_{j'}$,
where $g(0) \equiv g$, $g_j(0) = \gamma e^{-\frac{(j-l)^2}{2\sigma^2}+2\pi ik j}$, $j\in R_A$, and $g_j(0) = 0$ otherwise, and $\triangle$ is the $N\times N$ matrix whose matrix elements are given by $[\triangle]_{jj'} = \delta_{j,j'+1} + \delta_{j,j'-1}$, $1\le j,j'\le N$.

The operation $U_1$ corresponds to depositing a fermion into the lattice in a discretisation of a \emph{gaussian modulated} wavefunction. After some time $t$ the particle will have propagated out of the region $R_A$. Thus, Alice's region will be indistinguishable from the vacuum $|\Omega\rangle$. In this case the action of the next encoding operation $U_2$ will be proceed just as for $U_1$. However, the application of $U_2$ may not proceed due to some residual amplitude for the particle to remain in region $R_A$ and consequently to be swapped back to the ancilla $A_2$. Such a situation counts as an error. We now argue that the magnitude of this error is captured by the quantity 
\begin{equation}
	\|\{g(0), g^\dag(t)\}\| = |\langle g(0)|g(t)\rangle|,
\end{equation}
where $|g(t)\rangle = g(t)|\Omega\rangle = \sum_{j} g_j(t) |j\rangle$ is a single-particle state. Let's temporarily assume that this quantity is zero for all $t$. In this case there is no amplitude for a particle to be swapped back into the ancillas and the state of the system after $M$ messages are created is given by
\begin{multline}\label{eq:psim}
	|\Psi_M\rangle = U_M e^{-iHt}U_{M-1} \cdots U_2 e^{-iHt} U_1 |\psi_1\rangle \cdots |\psi_M\rangle |\Omega\rangle \\ = 
	|\mathbf{0}\rangle (c_M\mathbb{I} + d_Mg^\dag(0))\cdots (c_1\mathbb{I} + d_1g^\dag((M-1)t))|\Omega\rangle,
\end{multline} 
In this case we can intepret the state $|\Psi_M\rangle$ as that of $M$ free fermions moving independently without any interference effects whatsoever.

After a certain time $T$ has elapsed the first signal will have reached Bob's decoding region. Bob then decodes the message into his first ancilla by applying the operation
\begin{equation}
	V_\beta = e^{-i\frac{\pi}{2}(\sigma^{+}_\beta\sigma^{-}_\beta hh^{\dag} + \sigma^{-}_\beta\sigma^{+}_\beta h^{\dag}h)}e^{i\frac{\pi}{2} (\sigma^{+}_\beta h + \sigma^{-}_\beta h^{\dag})},
\end{equation}
where $\sigma^{\pm}_\beta$ acts on qubit $B_\beta$ and $h = \gamma'\sum_{j \in R_B} g_j(T) a_j,$
where $\gamma'$ is chosen to ensure the anticommutation relation $\{h,h^\dag\} = 1$. If we temporarily assume that $g(T)$ is given by 
$g(T) = \gamma\sum_{j\in R_B} g_j(T) a_j$ we see that $V_1$ is the optimal way to decode the signal as it completely swaps the fermion out of the lattice into the first ancilla. Evidently in the case where $|g(T)\rangle$ does not have this form then the decoding operation is not completely successful. Again a key role is played by the quantity 
\begin{equation}
	\|\{g(T), h^\dag\}\| = |\langle g(T)|h\rangle|,
\end{equation}
when it is equal to $1$ the decoding is completely successful. This quantity, when not unity, is interpreted as the discretisation error  $\epsilon_D$. 

Let's now consider the realistic situation where $\|\{g(0), g^\dag(t)\}\| \not=0$; it is clear that the system state is no longer of the form $|\Psi_M\rangle$ given by Eq.~(\ref{eq:psim}). In general we have that
\begin{equation}
	U_M e^{-iHt} \cdots U_2 e^{-iHt} U_1 |\psi_1\rangle \cdots |\psi_M\rangle |\Omega\rangle = |\Psi_M\rangle + |\Gamma_M\rangle,
\end{equation}
where now $|\Psi_M\rangle$ is of the form Eq.~(\ref{eq:psim}) (but no longer assumed normalised) and $|\Gamma_M\rangle$ is an orthogonal state which we interpret as an error. Using a combination of the triangle inequality, the leibniz property of the commutator, the anticommutation relations for $g$ and $g^\dag$, and the fact that $\|g\| \le 1$, we obtain the bound
\begin{equation}
	\epsilon_E = \||\Gamma_{M}\rangle\| \le 3\sum_{j=1}^{M-1} (M-j)|\langle g(0)|g(jt)\rangle|.
\end{equation}

where $\epsilon_E$ is the encoding error. Thus the errors occurring during the running of our protocol can be completely understood in terms of the overlaps $\langle g(0)|g(t)\rangle$ and $\langle g(t)|h\rangle$. Both of these quantities pertain to the single-particle sector; we have reduced our problem to understanding the dynamics of a single particle propagating through the lattice. This problem is now very well understood in the literature, and a comprehensive study may be found, for example, in \cite{osborne:2004a}. We summarise the pertinent results here (see also the supplementary material). Firstly, the gaussian modulated wavefunction $|g(0)\rangle$ with wavenumber $k$ propagates through the system at the group velocity $v(k) = \frac{d\omega(k)}{dk} = -\frac{4\pi}{N}\sin\left(\frac{2\pi}{N}k\right)$ through the system. Choosing $k_0= N/4$ maximises this group velocity and we obtain $v(k_0) = -4\pi/N$ (recall that Alice is located at physical position $0$ and Bob's at position $1/2$). Thus it takes a time $t = N/(8
\pi)$ for the wavepacket to travel from $R_A$ to $R_B$. Secondly, as long as the wavepacket is broad enough it retains its gaussian shape (up to some small errors) throughout the propagation from Alice's region to Bob's region. It turns out that an initial width $\sim N^{\frac13}$ is sufficient to ensure that the wavepacket doesn't disperse too quickly. This is because a wavepacket of physical width $L(0)$ corresponds, in momentum space, to a wavepacket of width $L^{-1}(0)$ centred on wavenumber $k = N/4$. Around this wavenumber the dispersion relation $\omega(k) = 2\cos(\frac{2\pi}{N}k)$ is approximately linear, with a small cubic correction. To estimate the broadening $L(t)$ due to dispersion coming from the cubic correction we use the \emph{third-order broadening factor} \cite{agrawal:2002a}
\begin{equation}\label{eq:3rdbraod}
\frac{L(t)}{L(0)} =
\left[1+\frac12\left(\frac{\omega'''(k_0)t}{\sqrt{2}L^3(0)}\right)^2\right]^{\frac12}.
\end{equation}
We solve this equation for $t=N/(8\pi)$ and find that a choice of $L(0) \sim N^{\frac13}$ is sufficient to ensure that the wavepacket spreads by at most a constant amount. Obviously there are errors coming from the truncation of the Alice's wavepacket to the region $R_A$, and from the discretisation. One can check that all these errors can be made exponentially small by redefining the constants $\sigma$ and $|R_A|$ (these calculations are presented in the supplementary material).

We find  $|\langle g(0)|g(t)\rangle| \sim e^{-\kappa^2 t^2 N^{-\frac23}},$ where $\kappa$ is a constant that only depends (logarithmically) on the desired errors $\epsilon_P$ and $\epsilon_E$, where $\epsilon_P$ is the propagation error. Thus, if we want to reduce the total encoding error arising from $\||\Gamma_{M}\rangle\|$ to $\epsilon_E$ it is sufficient to wait a time $t\sim \nu N^{\frac13}$, where the constant $\nu$ can be chosen to scale as $\nu\sim -\log(\epsilon)$. The decoding error experienced by Bob is proportional to $|\langle h|g(T)\rangle|$, where $T\sim N$ is the time taken for a wavepacket to enter Bob's region $R_B$. As has been argued previously \cite{osborne:2004a}, this quantity is simply related to the weight of the wavefunction in Bob's region. By choosing $|R_B|$ to scale as $N^{\frac13}$ sites this quantity can be increased towards $1$ exponentially quickly. 

The average fidelity $\mathcal{F}_\alpha$ for each of the decoded qubits $B_\alpha$ is thus straightforwardly bounded from below by $\mathcal{F}_\alpha \ge 1-\epsilon_E-\epsilon_P-\epsilon_D$. As we've argued, these three sources of error are, in turn, reduced exponentially fast in $c$ by redefining $\sigma$, $|R_A|$, and $|R_B|$ by a constant factor $c$.

It is inevitable that during the running of the protocol errors will build up in the system. Thus the achievable rate will also be implicitly determined by the maximum running time $T_{\text{max}}$ desired because the errors depend on $M$, the \emph{total} number of messages to be sent: the error estimate above only applies if $M \le N^{\frac23}$. In the case where $M = \lambda N^{\frac23}$ we need to add up the errors linearly in $\lambda > 1$ in the obvious way. Once the accumulated error goes above some prespecified threshold it is necessary to cool the system down again to the ground state. If we assume this takes some \emph{constant} time then we learn that the achievable rate is not affected.  

To conclude the description of our protocol we have to argue that Alice and Bob can carry out the encoding and decoding steps with access to only a single site. It turns out that a method to achieve this has already been described in \cite{haselgrove:2005a} in the spin-system setting. One can check that it applies with the obvious  modifications to the situation we have here.

Our protocol applies to the tight-binding model of electrons propagating around a ring. However it also extends without modification to the chain geometry. All that needs to be checked is that the relevant results concerning the free propagation of waves extend to the line-segment geometry. This is a straightforward exercise and is left to the reader. 

There are two natural extensions of our model. The first is to the setting where fermion-fermion interactions are included and the second is to the setting where signals are encoded in different initial wavepackets. In the first generalisation we model interactions via the (spin-less) $t-J$ model:
$H_{t-J} = -t\sum_{j=1}^{N-1} a_j^\dag a_{j+1} +\text{h.c.} + J\sum_{j=1}^{N-1} n_j n_{j+1}.$ (Our analysis also applies to the case with spin.)
In this case we may apply the previous analysis by noticing that  $\|e^{-iH_{t-J}s} |\Psi_M\rangle - e^{-iHs} |\Psi_M\rangle\| \le |s|\epsilon_{I}.$

This follows from triangle inequality and noting that the particle-particle interaction term $H_I = \sum_{j=1}^{N-1} n_j n_{j+1}$, when applied to $|\Psi_M\rangle$, satisfies $\|H_I|\Psi_M\rangle\| = \epsilon_I.$

The magnitude of $\epsilon_I$ may be reduced exponentially by linearly increasing $t$, the time between signals. This is because the state of the system is comprised of well-separated \emph{single particles}, and hence particle-particle interactions are negligible. A similar argument applies to the Hubbard model $H_{\text{HM}} = -t\sum_{j,\sigma} (c_{j,\sigma}^\dag c_{j+1,\sigma} + \text{h.c.}) + U\sum_{j=1}^N n_{j,\downarrow}n_{j,\uparrow},$ where $c_{j,\sigma}$ annihilates an electron with spin $\sigma$ at site $j$. In this case, our protocol proceeds as before after identifying $a_j$ with $c_{j,\downarrow}$ and ignoring the other spin degree of freedom. Because of the onsite interactions it is impossible to exploit the second spin degree of freedom to run two instances of the protocol in parallel without some nontrivial modification.

The second generalisation is for Alice to create additional signal fermions in wavepackets $\ell_k(x)$ \emph{orthogonal} to the gaussian wavepacket. Indeed, the \emph{hermite polynomials} naturally suggest themselves here. The nearly linear dispersion relation suggests that these orthogonal wavepackets also propagate without change of shape, and hence provide additional communication channels between Alice and Bob. However, a hermite polynomial is very close to a \emph{translation} of a gaussian. This implies that these wavepackets propagate at a different group velocity. If one attempts to find $N^{1/3}$ such orthogonal wavepackets one runs into the problem that the approximations involved in the derivation of the group velocity and dispersion relation begin to break down, especially at the higher frequencies. This generalisation may provide a \emph{constant} qubit rate but is unlikely to work for many realistic models (except for the tight-binding model) due to the fermion-fermion interactions.

We have described a protocol whereby Alice and Bob, having access to only small regions of a system of electrons moving in the presence of a lattice of binding atoms, can communicate quantum information with arbitrarily high fidelity at a rate of $N^{-\frac13}$ qubits per unit time. This result improves considerably upon the extant protocols for quantum information wires in two important ways. Firstly, the rate we achieve here is much greater than that encountered in the quantum spin-system case, and improves on the rate achieved by \cite{kay:2010a} for a specially engineered chain. Secondly, our protocol applies to naturally occurring systems modelled by the tight-binding, $t-J$, and Hubbard models. There is considerable scope for further work following our contribution: the extension of our analysis to the bosonic case, e.g., to the Bose-Hubbard model modelling cold atoms in optical lattices, certainly merits investigation. Also, a more detailed analysis of the case where different encoding wavefunctions may yield better --- possibly even \emph{constant} --- qubit rates.

\begin{acknowledgments}

This paper originated from conversations with Noah Linden whose support and input is sincerely and gratefully acknowledged.

We would like to thank Daniel Burgarth, Matthias Christandl,
Nilanjana Datta, Artur Ekert, Henry Haselgrove, Simone Severini,
and Andreas Winter for many helpful discussions. We are grateful
to the EU for support for this research under the IST project
RESQ.
\end{acknowledgments}

\onecolumngrid
\appendix
\section{Supplementary material}

\subsection{Bounding the error terms}
We bound the error term $|\Gamma_M\rangle$ as follows: using the identity Eq.~(\ref{eq:Uidentity}) we have that
\begin{equation}
	U_{M} e^{-iHt} \cdots e^{-iHt} U_1 |\psi_1\rangle \cdots |\psi_{M}\rangle |\Omega\rangle 
	= |\Psi_{M}\rangle  + |\Gamma_{M}\rangle,
\end{equation}
where
\begin{equation}
	|\Gamma_{M+1}\rangle = |\Delta_1\rangle + |\Delta_2\rangle +|\Delta_3\rangle+ |\Delta_4\rangle,
\end{equation}
and
\begin{equation}
	\begin{split}
		|\Delta_1\rangle &= U_{M}|\psi_{M+1}\rangle e^{-iHt}|\Gamma_{M-1}\rangle,\\ 
		|\Delta_2\rangle &= d_{M}|1\rangle (\mathbb{I}-g(0)g^\dag(0))e^{-iHt}|\Psi_{M-1}\rangle,\\		
		|\Delta_3\rangle &= - c_{M}|0\rangle g^{\dag}(0)g(0) e^{-iHt}|\Psi_{M-1}\rangle,\\
		|\Delta_4\rangle &=  c_{M}|1\rangle g(0) e^{-iHt}|\Psi_{M-1}\rangle.
	\end{split}
\end{equation}
The norms of each of these four terms can be bounded as follows. Firstly we have by unitary invariance of the norm that $\||\Delta_1\rangle\| \le \||\Gamma_{M-1}\rangle\|$.
The second may be bounded by noticing that
\begin{equation}
	\begin{split}
		\||\Delta_2\rangle\| &\le |d_{M}| \|[ g(0)g^\dag(0),   (c_{M-1}\mathbb{I} + d_{M-1}g^\dag(1))\cdots (c_1\mathbb{I} + d_1g^\dag((M-1)t))]|\Omega\rangle\| \\
		&\le \sum_{j=1}^{M-1} |d_j|\|(c_{M-1}\mathbb{I} + d_{M-1}g^\dag(1))\cdots (c_{j+1}\mathbb{I} + d_{j+1}g^\dag((M-j-1)t)) [g(0)g^\dag(0), g^\dag((M-j)t)]\times \\ &\quad\quad\quad (c_{j-1}\mathbb{I} + d_{j-1}g^\dag((M-j+1)t)) \cdots (c_1\mathbb{I} + d_1g^\dag((M-1)t))|\Omega\rangle\| \\
		&\le \sum_{j=1}^{M-1} \|\{g(0), g^\dag((M-j)t)\}\|\|(c_{M-1}\mathbb{I} + d_{M-1}g^\dag(1))\cdots (c_{j+1}\mathbb{I} + d_{j+1}g^\dag((M-j-1)t)) g^\dag(0) \times \\ &\quad\quad\quad (c_{j-1}\mathbb{I} + d_{j-1}g^\dag((M-j+1)t)) \cdots (c_1\mathbb{I} + d_1g^\dag((M-1)t))|\Omega\rangle\| \\
		&\le \sum_{j=1}^{M-1} \|\{g(0), g^\dag((M-j)t)\}\|\|g^\dag(0)\|\|(c_{M-1}\mathbb{I} - d_{M-1}g^\dag(1))\cdots (c_{j+1}\mathbb{I} - d_{j+1}g^\dag((M-j-1)t)) \times\\ &\quad\quad\quad (c_{j-1}\mathbb{I} + d_{j-1}g^\dag((M-j+1)t)) \cdots (c_1\mathbb{I} + d_1g^\dag((M-1)t))|\Omega\rangle\| \\
		&\le  \sum_{j=1}^{M-1} \|\{g(0), g^\dag(jt)\}\|,
	\end{split}
\end{equation}
where we've used the triangle inequality, the leibniz property of the commutator, the fact that $\|g\| \le 1$, and that $\| (c_M\mathbb{I} + d_Mg^\dag(1))\cdots (c_2\mathbb{I} + d_1g^\dag(Mt))|\Omega\rangle\| \le 1$ for all $c_j$ and $d_j$. The third term is bounded using exactly the same argument, and we obtain
\begin{equation}
	\||\Delta_3\rangle\| \le \sum_{j=1}^{M-1} |\langle g(0)| g(jt)\rangle |.
\end{equation}
The final term is bounded using anticommutation relations for $g(0)$ and $g(t)$:
\begin{multline}
	g(0) e^{-iHt}|\Psi_{M-1}\rangle = \sum_{j=1}^{M-1} d_{j} \langle g(0)|g((M-j)t)\rangle (c_{M-1}\mathbb{I} - d_{M-1}g^\dag(1))\cdots(c_{j+1}\mathbb{I} - d_{j+1}g^\dag((M-j-1)t)) \times\\  (c_{j-1}\mathbb{I} + d_{j-1}g^\dag((M-j+1)t)) \cdots (c_1\mathbb{I} + d_1g^\dag((M-1)t))|\Omega\rangle.
\end{multline}
Taking the norm gives us, via the triangle inequality,
\begin{equation}
	\||\Delta_4\rangle\| \le \sum_{j=1}^{M-1} |\langle g(0)| g(jt)\rangle |.
\end{equation}

Thus we learn that
\begin{equation}
	\||\Gamma_{M}\rangle\| \le \||\Gamma_{M-1}\rangle\| + 3 \sum_{j=1}^{M-1} |\langle g(0)| g(jt)\rangle| \le 3\sum_{j=1}^{M-1} (M-j)|\langle g(0)|g(jt)\rangle|.
\end{equation}

\section{Dynamics in the single-particle sector}
In this appendix we review some results concerning the propagation of discrete gaussian wavepackets for scalar particles propagating in a regular one-dimensional lattice. 

Throughout this appendix we work in the \emph{single-particle subspace} $\mathcal{H}_S$. This is the subspace of hilbert space with basis $|j\rangle \equiv a_j^\dag|\Omega\rangle$, $j = 1, 2, \ldots, N$. Since the tight-binding model Eq.~(\ref{eq:tightbinding}) preserves particle number then if the system begins in $\mathcal{H}_S$ it remains there for all time. The first thing to note is that the matrix-elements of Eq.~(\ref{eq:tightbinding}) in $\mathcal{H}_S$ are given by
\begin{equation}
	[\triangle]_{jk} = \langle j|H|k\rangle.
\end{equation}
where $\triangle$ is the $N\times N$ matrix whose matrix elements are given by $[\triangle]_{jj'} = \delta_{j,j'+1} + \delta_{j,j'-1}$, $1\le j,j'\le N$, where we identify the first site with the $(N+1)$th site: $1\equiv N+1$. 

Since $\triangle$ is circulant it may be diagonalised via a discrete fourier transform; we obtain
\begin{equation}
	\triangle = \sum_{k=1}^{N} \omega(k) |W(k)\rangle\langle W(k)|,
\end{equation}
where 
\begin{equation}
	\omega(k) = 2\cos\left(\frac{2\pi}{N}k\right)
\end{equation}
is the \emph{dispersion relation} with corresponding eigenvectors 
\begin{equation}
	|W(k)\rangle \equiv \frac{1}{\sqrt{N}}\sum_{j=1}^N \mu^{jk}|j\rangle,
\end{equation}
and where $\mu$ is the $N$th root of unity, $\mu=e^{\frac{2\pi}{N}i}$. 

As noted in the body of the paper, the operator $\triangle$ may be understood as the discretised kinetic energy operator. Hence, the dynamics generated by $\triangle$ are physically equivalent to that of a freely propagating particle. These are now familiar facts and, with the appropriate qualifications, may be made mathematically rigourous (see, e.g., any book on the numerical analysis of PDE, say). For simplicity, rather than summarise the relevant (large) mathematically rigourous literature, we content ourselves here with a discussion at the level of physical rigour. The reader may be assured that all statements here can be brought to the level of mathematical rigour with a suitable amount of technical effort. 

The first step is extend, via the natural map, every state $|\phi\rangle = \sum_{k=1}^N \widehat{\phi}(k) |W(k)\rangle$ of our discrete system to a function $\phi:\mathbb{S}^1\rightarrow \mathbb{C}$ in $L_2(\mathbb{S}^1)$:
\begin{equation}
	|\phi\rangle \mapsto \phi(x) = \sum_{k=1}^N \widehat{\phi}(k) e_k(x),
\end{equation}
where $e_k(x) = e^{2\pi i kx}$ and $x\in [0,1)$. Note that $\phi$ is normalised with respect to the standard $L_2$ inner product. In this way we see that 
\begin{equation}
	\langle j| \phi\rangle = \phi(x_j),
\end{equation}
where $x_j = j/N$. Recall that the momentum operator $\widehat{p} = -i\frac{d}{dx}$ acts on $\phi$ as a multiplication operator
\begin{equation}
	\widehat{p} \phi = 2\pi \sum_{k=1}^N k \widehat{\phi}(k) e_k(x).
\end{equation} 
Now the image $\widehat{\triangle}$ of the operator $\triangle$ acts as the multiplication operator  
\begin{equation}
	\widehat{\triangle} \phi = \sum_{k=1}^N \omega(k)\widehat{\phi}(k) e_k(x),
\end{equation} 
where $\omega(k) = 2\cos\left(\frac{2\pi}{N}k\right)$. We now focus on momenta $k$ near $k_0 = \frac{N}{4}$ (assuming $4|N$) to maximise the group velocity: suppose that
\begin{equation}
	k = k_0 + l,
\end{equation}
with $l$ small in comparison to $k_0$. We then expand the dispersion relation $\omega(k)$
\begin{equation}
	\begin{split}
		\omega(k) &= \omega(k_0) + l \frac{d\omega}{dk}(k_0) + \frac{l^2}{2!}\frac{d^2\omega}{dk^2}(k_0) + \frac{l^3}{3!}\frac{d^3\omega}{dk^3}(k_0) + \cdots \\
		&= 0 - \frac{4\pi}{N}l + 0 + \frac{2}{3!}\frac{(2\pi)^3}{N^3}l^3 + \cdots.
	\end{split}
\end{equation}
Now suppose that
\begin{equation}
	\phi(x) = \int_{-\Lambda}^\Lambda \widehat{\phi}(l) e_{k_0+l}(x) \, dl = e_{k_0}(x)\int_{-\Lambda}^\Lambda \widehat{\phi}(l) e_{l}(x) \, dl = e_{k_0}(x)\phi_0(x),
\end{equation}
where $\Lambda$ is a cutoff.
Clearly, for $\Lambda \sim o(N)$, we have that the action of $\widehat{\triangle}$ on $\phi$ is well approximated by the action of the wave operator  
\begin{equation}
	\widehat{\triangle} \sim -\frac{2}{N}\widehat{p} + \frac{2}{3!N^3}\widehat{p}^3  
\end{equation}
on $\phi_0(x)$, i.e.,
\begin{equation}
	\widehat{\triangle} \phi(x) = e_{k_0}(x) \left(-\frac{2}{N}\widehat{p} + \frac{2}{3!N^3}\widehat{p}^3\right)\phi_0(x) + \cdots.
\end{equation}
In particular, with the choice $\Lambda = \kappa N^{\frac{2}{3}}$, we have that the action of $\widehat{\triangle}$ on $\phi_0(x)$ is equivalent to  
\begin{equation}
	\widehat{\triangle}\phi_0(x) \sim \left(-\frac{2}{N}\widehat{p} + \frac{2}{3!N^3}\widehat{p}^3\right)\phi_0(x) + O(N^{-\frac{5}{3}}).
\end{equation}
Thus, as long as the discrete fourier representation of $|\phi\rangle$ obeys
\begin{equation}\label{eq:cutoffdiscfour}
	|\phi\rangle = \sum_{k = 1}^N \phi(k)|W(k)\rangle  = \sum_{k = k_0 - \Lambda}^{k_0+\Lambda} \phi(k)|W(k)\rangle,
\end{equation}
we can approximate the dynamics of $\triangle$ via the dynamics of the wave operator $-\frac{2}{N}\widehat{p} + \frac{2}{3!N^3}\widehat{p}^3$. Evidently this approximation will only hold for $t \sim o(N^{\frac{5}{3}})$. However, this is more than enough for our situation as $t\sim N$ is the time it takes for the wavepacket to traverse the distance between Alice and Bob. 

Now we investigate the smallest width in real space a wavepacket obeying the constraint $\Lambda = \kappa N^{\frac{2}{3}}$ can have. To do this we introduce the dirac comb
\begin{equation}
	\eta_{N}(x) = \frac{1}{\sqrt{N}} \sum_{j\in \mathbb{Z}} \delta(x-j/N).
\end{equation}
Suppose that there is a $L_2(\mathbb{S}^1)$ function $\phi(x)$ such that $\phi(x) = 0$ for $x\not\in (0,1)$ and $\langle j|\phi\rangle = \phi(x_j)$ (indeed, such a function always exists, and further, $\phi$ may be chosen to be $C^\infty$). In this case we have that
\begin{equation}
	\mathcal{F}[\phi(x)\eta_N(x)](k) = \frac{1}{\sqrt{N}}\sum_{j=1}^N e^{\frac{2\pi}{N}i jk} \langle j|\phi\rangle, 
\end{equation}
for $k = 1, 2, \ldots, N$, 
where 
\begin{equation}
	\mathcal{F}[f(x)](k) = \int_{-\infty}^{\infty} e^{2\pi i kx} f(x)\, dx, 
\end{equation}
is the fourier transform operator. In this way we see that
\begin{equation}
	|\phi\rangle = \sum_{k=1}^{N} (\widehat{\phi}\star \widehat{\eta})(k) |W(k)\rangle.
\end{equation}
Now we recall that a gaussian
\begin{equation}
	g(x) = \frac{e^{-\frac{(x-x_0)^2}{2\sigma^2}}}{\sqrt{\sigma}\pi^{\frac14}}
\end{equation}
has fourier transform
\begin{equation}
	\widehat{g}(k) = \pi^{\frac14}\sqrt{2\sigma} e^{2\pi i k x_0} e^{-2\pi^2 \sigma^2 k^2}.
\end{equation}
The objective is to choose $\phi(x) = g(x)$. Unfortunately such a choice doesn't satisfy $\phi(x) = 0$ for $x\not\in (0,1)$. (And, indeed, such a wavefunction is nontrivial outside Alice's region.) However, we can choose $\sigma$ so that this constraint is satisfied up to an exponentially small error. 

Let's now calculate the $\sigma$ required to \emph{approximately} satisfy the constraint $\phi(x) = 0$ for $x\not\in R_A$ \emph{and} the constraint $\widehat{\phi}_0(k)= 0$ for $k \not\in [-\Lambda, \Lambda]$ required by Eq.~(\ref{eq:cutoffdiscfour}). The second constraint is satisfied up to an exponentially small error $e^{-c}$ with the choice
\begin{equation}
	\sigma^2 = \frac{c}{2\pi^2\kappa^2 N^{\frac{4}{3}}}.
\end{equation}
This corresponds, in real space, to a gaussian $g(x)$ with characteristic width 
\begin{equation}
	L(0) = \frac{N^{-\frac{2}{3}} \sqrt{c}}{2\pi \kappa}.
\end{equation}
This corresponds, in lattice units, to a region $R_A$ of width $|R_A| = NL(0) = \frac{N^{\frac{1}{3}} \sqrt{c}}{2\pi \kappa}$ sites.

Now, under the dynamics generated by $\triangle$ a gaussian wavefunction does not remain a gaussian. Instead, it
becomes an Airy function \cite{agrawal:2002a}. To estimate the width of the evolved gaussian $|g(0)\rangle = e^{-it\triangle}|g(0)\rangle$ (which is found from $g(x,t) = e^{-it\widehat{\triangle}}g(x,0)$) we need to
use the \emph{third-order broadening factor} \cite{agrawal:2002a}
\begin{equation}
\frac{L(t)}{L(0)} =
\left[1+\frac12\left(\frac{\omega'''(k_0)t}{\sqrt{2}L^3(0)}\right)^2\right]^{\frac12}.
\end{equation}
We solve this equation for $t=N$ (the time it takes to traverse the lattice): we find that with our choice of 
that $L(0)$ such a wavepacket spreads by at most a constant amount.

The third-order broadening factor may derived by estimating the overlap $|\langle g(0)|g(t)\rangle|$ for arbitrary $t$. Using the fact that
\begin{equation}
	\langle g(0)|g(t)\rangle = ( g(0), e^{-it\widehat{\triangle}}g(0))
\end{equation}
we reduce our problem, via Parseval's relation, to the estimation of the Fourier-Airy integral:
\begin{equation}
	2\sigma\sqrt{\pi} \int_{-\infty}^\infty e^{-4\pi^2 \sigma^2 k^2} e^{\frac{4\pi i}{N}tk - \frac{2i}{3!}\frac{(2\pi)^3}{N^3}tk^3} \, dk. 
\end{equation}
Choosing $t = \frac12 x_1 N^{\frac13}$ (the time it takes for a wavepacket to propagate from Alice's region) we have that
\begin{equation}
	2\sigma\sqrt{\pi} \int_{-\infty}^\infty e^{-4\pi^2 \sigma^2 k^2} e^{2\pi i N^{-\frac23} x_1 k - \frac{i}{3!}(2\pi)^3 N^{-\frac{8}{3}} x_1 k^3} \, dk. 
\end{equation}
Notice that, because of the gaussian weighting, the integrand is nontrivial only for $k \lesssim \kappa N^{\frac{2}{3}}$, so that changing variables to $k = \kappa N^{\frac23} l$ we have
\begin{equation}
	\frac{\sqrt{2c}}{\sqrt{\pi}} \int_{-\infty}^\infty e^{2\pi i \kappa  l  x_1} e^{-2c l^2} e^{- \frac{i}{3!}(2\pi)^3 N^{-\frac{2}{3}} x_1 \kappa^3 l^3} \, dl. 
\end{equation}
We now expand in the smallness of the third exponent:
\begin{equation}
	\frac{\sqrt{2c}}{\sqrt{\pi}}\int_{-\infty}^\infty e^{- 2cl^2 } e^{2\pi i \kappa  l  x_1}\left(1 - \frac{i}{3!}(2\pi)^3 N^{-\frac{2}{3}} x_1 \kappa^3 l^3+ \cdots\right) \, dk.
\end{equation}
The dominant term is evidently the first term in the expansion (this is only valid as long as $x_1$ is $O(1)$, i.e., on timescales longer than the lattice scale; for very short times $x_1 \sim N^{-1}$ the expansion breaks down):
\begin{equation}
	e^{-\frac{\pi^2\kappa^2 x_1^2}{2c}}.
\end{equation}
Higher order terms come from differentiation of $e^{-\frac{\pi^2\kappa^2 x_1^2}{2c}}$ with respect to $\kappa$, and the remainder $R(\kappa)$ is easily estimated to scale as
\begin{equation}
	|R(\kappa)| = \text{const.}\times e^{-\frac{\pi^2\kappa^2 x_1^2}{2c}},
\end{equation}
where the constant depends only on $c$ and $\kappa$. Recall that $x_0 \le x_1 \le N^{\frac23}$, where $x_0$ corresponds to the centre of Alice's region and $x_1 = N^{\frac23}/2$ corresponds to a separated physical position. By increasing $x_1$ we can reduce this overlap to zero doubly exponentially fast. Hence we have our estimate
\begin{equation}
	|\langle g(0)|g(t)\rangle| \sim \text{const.}\times e^{-\frac{\pi^2\kappa^2 x_1^2}{2c}}.
\end{equation} 
This tells us that to reduce the total error 
\begin{equation}
	\||\Gamma_M\rangle\| \lesssim |\langle g(0)|g(t)\rangle |
\end{equation}
to less than $\epsilon$ we need only wait a time $t \sim \times N^{\frac13}$ between signals.

\end{document}